\documentclass[aps,english,showpacs,twocolumn,pra]{revtex4-1}
\usepackage{amsfonts}
\usepackage{amssymb}
\usepackage{amsmath}
\usepackage{graphicx}
\usepackage{epsfig}
\usepackage{makecell}
\usepackage{enumerate}
\usepackage{mathrsfs}
\usepackage{color,xcolor}

\begin{document}

\title{Time-reversal symmetric topological metal}
\author{L. C. Xie}
\author{H. C. Wu}
\author{L. Jin}
\email{jinliang@nankai.edu.cn}
\author{Z. Song}

\begin{abstract}
Topological metals possess gapless band structures accompanied with
nontrivial edge states. Topological metals are created from the topological
insulators by adjusting the magnetic flux. Here we propose the time-reversal
symmetry protected topological metal without employing the magnetic flux.
Topological metallic phase presents although the Chern number vanishes. In
the topologically nontrivial phase, three different cases in the metallic
phase are distinguished from the band-touching and in-gap features of the
topological edge states. These findings shed light on the time-reversal
symmetric topological metals and greatly simplify the realization of
topological metals.
\end{abstract}

\affiliation{School of Physics, Nankai University, Tianjin 300071, China}
\maketitle

\section{Introduction}

The introduction of topology in the condensed matter physics greatly changes
the traditional way of classifying phases of matter, novel topological
states including insulators, superconductors, semimetals, and metals are
discovered \cite%
{E.Mele2007,Kane2010,XGWen2017,T.Das2016,SCZhang2011,L.Balents2011,AMRappe2012,A.Vish2018,AB16,BYan17,CKChiu16,ABansil16,LJin19,Kawabata19,Ashida20}%
. These topological phases all support robust edge states, which are counted
by the corresponding topological invariant that capturing the topology of
the bulk band \cite{DJTh1982,DXiao2010,Kane2010b}. The tremendous progresses
in various platforms of photonics, acoustics, and electronic circuits allow
us to study these topological phases in experiments, which provide crucial
enlightenments for topological materials \cite%
{YRan2009,LJLang2012,XJLiu13,MAide13,LLu14,SYYu18,YBYang18,SLZhu19,XDChen19,YXu19,SImh18,THel19,YLu19}%
.

Topological insulators have a fully gapped bulk band structure like ordinary
insulators, but topological insulators support conductive edge states on the
surface of the sample, where electrons move along the surface of materials 
\cite{Kane2010}. The Chern insulator is a typical two-dimensional (2D)
topological insulator, which supports robust chiral edge states that are
predicted from the nonzero Chern number. A landmark development of
topological insulators is the discovery of time-reversal invariant
topological insulators \cite{S.M04,CLKane05a,BAB06,XLQi06}, where the Chern
number vanishes and becomes invalid for topological characterization;
alternatively, the $Z_{2}$ topological number classifies the time-reversal
invariant insulators \cite{CLKane05b} and these topological insulators are
featured by their helical edge states, being confirmed in nonlocal transport
experiment \cite{ARoth09}. The helical edge states have opposite spins and
counter-propagate on the boundaries in the opposite directions. The spin
Chern number \cite{LSheng05,EProdan09,YYang11}, valley Chern number \cite%
{FZhang11,FZhang13,MEzawa13}, and wave polarization \cite%
{RResta94,PDelp11,FLiu17} are developed to characterize the quantum spin
Hall effect.

Topological semimetals and metals are metallic states with nontrivial
gapless band structure and unique quasi-particle excitations \cite%
{A.Vish2018}, represented by Weyl and Dirac semimetals. In the Weyl (Dirac)
semimetals, energy bands are double (fourfold) degenerately connected by the
Weyl (Dirac) points \cite{SMura17,XWan11,ZWang13}. In addition, the
existence of other multiple degeneracy points has been discovered, like
threefold and sixfold \cite{ZZhu16,BJW16,Hweng16,SSingh18,GShan19,RChapai19}%
. For systems protected by certain symmetries, the band connection is a
closed node-line instead of nodes. The corresponding topological node-line
semimetals have been reported in the graphene network \cite%
{AAB11,RYu15,HMW15}.

Topological metals (TMs) are gapless phases featured by their separable
conduction and valence bands. Thus, the topological characterization for the
metals is similar as that for the insulators. TMs can be divided into
nodal-point, nodal-line, and nodal-surface TMs according to the dimension of
band crossing \cite{WWu18}; or alternatively be divided into type I and type
II TMs according to the dispersion close to the band crossing \cite%
{AAS15,YXu15,SLi17}. TMs in different spatial dimensions have been
extensively studied \cite{MBahari19,SYXu15,ZJWang16}. To create 2D
topological metal, the traditional method is through adjusting the magnetic
flux to tune the energy bands of the topological insulators \cite{XZYing18}.
However, introducing and tuning the magnetic flux is difficult in many
experimental platforms. Here we demonstrate a time-reversal symmetric
topological metal without the magnetic flux.

In this paper, we report the time-reversal symmetric topological metal in a
2D square lattice, which belongs to class AI in the tenfold way of
topological classification. The 2D square lattice possesses nontrivial
topology although the Chern number is zero ensured by the time-reversal
symmetry. The proposed topological metal is free from the magnetic flux and
significantly differs from the traditional topological metal created through
using the magnetic flux. The reflection symmetry ensures that the quantized
polarization; the zero polarization predicts the trivial topology in the
vertical direction. The Zak phase at the phase transition point
characterizes the topology in the horizontal direction. We demonstrate three
different cases in the topologically nontrivial phases that are
distinguished from their different types of topological edge states. The
topological phases as well as the topological edge states can be deformed
into each other without undergoing topological phase transition. Our
findings provide novel insights into time-reversal symmetric topological
metals.

The remainder of the paper is organized as follows. In Sec.~\ref{II}, the
time-reversal symmetric 2D square lattice is introduced. Section \ref{III}
presents the phase diagram and the energy band of the topological metallic
phase. Three different topological edge states are elucidated. Section~\ref%
{IV} provides the topological characterization. The discussion and
conclusion are summarized in Secs.~\ref{V} and~\ref{VI}.

\section{Time-reversal symmetric 2D square lattice}

\label{II}

\begin{figure}[tb]
\includegraphics[bb=55 455 365 755, width=7.8 cm,clip]{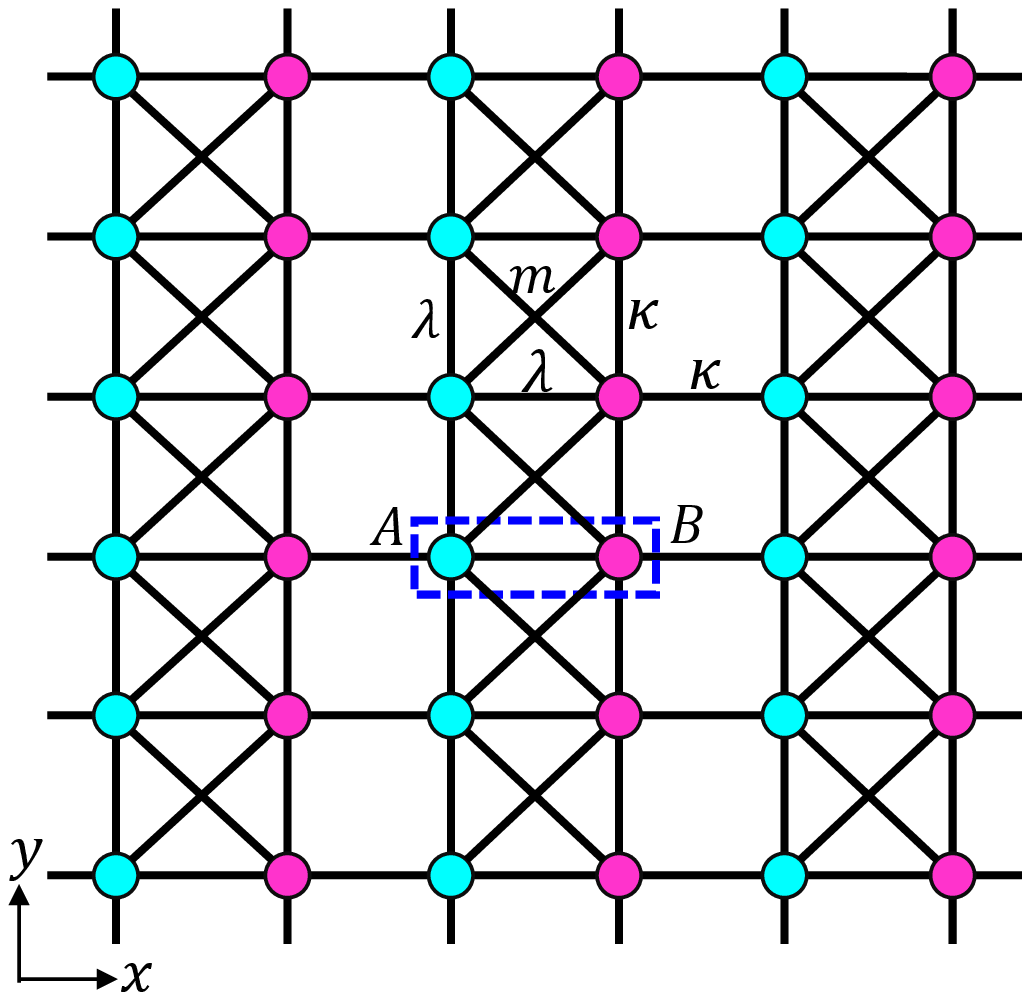}
\caption{Schematic of the 2D square lattice. The unit cell is consist of two
sublattices $A$ and $B$ in cyan and magenta as indicated inside the dashed
blue rectangle. $\protect\lambda$, $\protect\kappa$, and $m$ are the
couplings between the sublattices.}
\label{Lattice}
\end{figure}

We consider a time-reversal symmetric 2D square lattice. The schematic of
the lattice is illustrated in Fig.~\ref{Lattice}. The Hamiltonian of the 2D
square lattice in the real space is written in the form of%
\begin{eqnarray}
H &=&\sum\limits_{l,j}(\lambda a_{l,j}^{\dagger }b_{l,j}+\kappa
a_{l,j}^{\dagger }b_{l-1,j}+ma_{l,j}^{\dagger }b_{l,j+1}+  \notag \\
&&ma_{l,j}^{\dagger }b_{l,j-1}+\text{H.c.})+\lambda a_{l,j}^{\dagger
}(a_{l,j+1}+a_{l,j-1})  \notag \\
&&+\kappa b_{l,j}^{\dagger }(b_{l,j+1}+b_{l,j-1}),
\end{eqnarray}%
where $a^{\dagger }$, $b^{\dagger }$ $(a$, $b)$ represent the creation
(annihilation) operators for the sublattices $A$ and $B$. The dashed box in
blue indicates the unit cell of the system. The couplings $\lambda $ and $%
\kappa $ indicate the nearest neighbor coupling strengths, the coupling $m$
is the next nearest neighbor coupling strength. The 2D square lattice does
not enclose any magnetic flux and can be realized in many versatile
experimental platforms for the topological insulators including the
ultracold atoms in the optical lattice, the photonic crystals of coupled
optical waveguides/resonators, the acoustic metamaterial dubbed sonic
crystals, and the electric circuits \cite%
{MAide15,SMi18,THofma19,CHe16,ZJYang15,YGPeng16,JYLu18}.

Applying the Fourier transformation $\rho _{l,j}=M^{-1}\sum_{\mathbf{k}}e^{i%
\mathbf{k\cdot r}}\rho _{\mathbf{k}}$ $(\rho =a,b)$ to the two sublattices,
we rewrite the lattice Hamiltonian $H$ in the momentum space in the form of 
\begin{equation}
H=\sum_{\mathbf{k}}H\left( \mathbf{k}\right) =\sum_{\mathbf{k}}\psi _{%
\mathbf{k}}^{\dagger }h\left( \mathbf{k}\right) \psi _{\mathbf{k}},
\end{equation}%
where the basis is $\psi _{\mathbf{k}}=(a_{\mathbf{k}},b_{\mathbf{k}})^{T}$
and the Bloch Hamiltonian $h\left( \mathbf{k}\right) $ is a $2\times 2$
matrix%
\begin{equation}
h\left( \mathbf{k}\right) =d_{0}\sigma _{0}+\mathbf{d\cdot \sigma ,}
\label{H}
\end{equation}%
where $\sigma _{0}$ is the $2\times 2$ identical matrix, $d_{0}=(\lambda
+\kappa )\cos k_{y}$. $\mathbf{\sigma =(\sigma }_{x},\mathbf{\sigma }_{y},%
\mathbf{\sigma }_{z}\mathbf{)}$ is the Pauli matrices, and the effective
magnetic field $\mathbf{d}$ is%
\begin{eqnarray}
d_{x} &=&\lambda +2m\cos k_{y}+\kappa \cos k_{x},  \notag \\
d_{y} &=&\kappa \sin k_{x},  \label{B} \\
d_{z} &=&(\lambda -\kappa )\cos k_{y}.  \notag
\end{eqnarray}%
The energies of the Bloch bands are%
\begin{equation}
E_{\pm }=d_{0}\pm \sqrt{d_{x}^{2}+d_{y}^{2}+d_{z}^{2}},  \label{Energy}
\end{equation}%
where the subscripts $+$ and $-$ of energy represent the upper and lower
bands, respectively. Without loss of generality, we fix the coupling $m$ as
unity. Notably, the band gap closes at $\lambda =\pm \kappa $, associated
with the appearance of degenerate points (DPs) or degenerate lines.

The Bloch Hamiltonian $h\left( \mathbf{k}\right) $ of the square lattice is
protected by the time-reversal symmetry. Thus, the band energies satisfy $%
E_{\pm ,\mathbf{k}}=E_{\pm ,-\mathbf{k}}^{\ast }$ as reflected from Eq. (\ref%
{Energy}). The symmetries are important \cite{CKChiu16,Kawabata19,LJin21}.
Notably, the time-reversal symmetry ensures a zero Chern number for the
energy band; however, the topologically nontrivial phase still exists in the
2D lattice \cite{FLiu17,HCWu20}. $d_{0}$\ adjusts the energy band and
changes the band structure, and plays a crucial role in creating the
metallic phase when the strengths of the couplings $\lambda $ and $\kappa $
are not equal $\lambda ^{2}\neq \kappa ^{2}$. In the absence of $d_{0}$, the
bands are fully gapped except for the band touching and the system is in an
insulator phase.

The energy bands have highly degenerate points, where the energy bands
become fully flat. The highly degenerate points appear at $k_{y}=\pm \arccos
(-\lambda /2)$, but disappear as $\left\vert \lambda \right\vert >2$.
Time-reversal symmetry ensures that the two highly degenerate points are
symmetric about $k_{y}=0$. The upper and lower bands reduce to two momentum
independent flat bands%
\begin{equation}
\varepsilon _{\pm }=-(\lambda ^{2}+\kappa \lambda \pm \sqrt{\Delta })/2,
\end{equation}%
where $\Delta =4\kappa ^{2}+\kappa ^{2}\lambda ^{2}-2\kappa \lambda
^{3}+\lambda ^{4}$. Notice that the presence of $d_{0}$ breaks the symmetry
of the upper and lower bands about zero energy.

We emphasize that the proposed topological metal significantly differs from
the traditional topological metal created through altering the magnetic
flux. Their symmetry protections, topological classifications, and
topological invariants are different. The proposed topological metal without
magnetic flux is time-reversal symmetric and belongs to the symmetry class
AI \cite{SRu10}. In contrast, the topological metal with magnetic flux
breaks the time-reversal symmetry; for example, the topological metal in
Ref.~\cite{XZYing18} created via tuning the magnetic flux from the ordinary
metal is particle-hole symmetric and belongs to the symmetry class D.
Consequently, the topological invariants are different. In addition, the
energy bands are symmetric about $k_{y}=0$ in the time-reversal symmetric
topological metal. Thus, the edge states cross twice at zero energy as a
consequence of zero Chern number $C=0$; and the dispersions and velocities
near the zero edge states on the two sides of $k_{y}=0$ are opposite. We can
observe these features in the following section. However, the edge states
cross once at zero energy in the traditional topological metal without the
time-reversal symmetry protection as a consequence of nonzero Chern number $%
C=\pm1$.

\section{Metal-insulator transition}

\label{III}

\begin{figure}[tb]
\includegraphics[bb=0 0 560 550, width=7.8cm, clip]{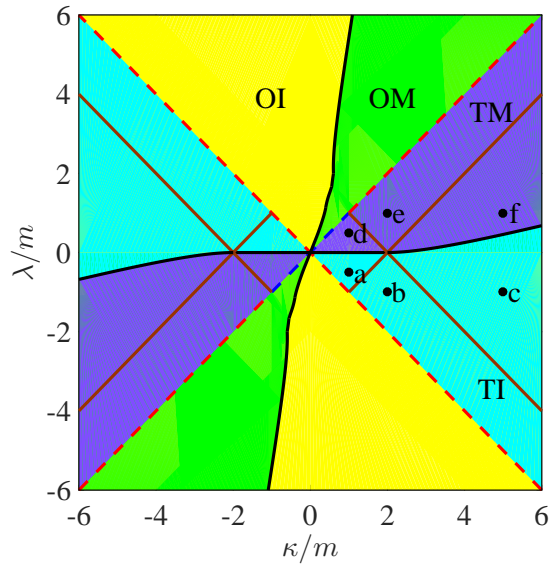}
\caption{Phase diagram of $h\left(\mathbf{k}\right)$ in the $\protect\kappa$-%
$\protect\lambda$ parameter space. Regions with different colors indicate
different phases, topological metal (TM) in purple; ordinary metal (OM) in
green; topological insulator (TI) in cyan; ordinary insulator (OI) in
yellow. The black solid lines indicate the Wilson transitions. The red
dashed lines indicate the topological phase transition with two DPs, while
the blue dashed line indicates the topological phase transition with four
DPs. The brown solid lines are the boundary that characterizes the
positional relationship between the edge states and bulk bands, indicating
three different cases in the topological metallic or insulator phase.}
\label{PhaseD}
\end{figure}

In this section, we discuss the topological phases. The prominent feature of
the considered 2D square lattice is the presence of topological metallic
phase. For a two-band model, if the bulk bands are separable and the Fermi
level lies in the band gap, the system is an insulator. Once the highest
energy level of the valence band exceeds the lowest energy level of the
conduction band, the two band energies partly overlap each other and the
Fermi level crosses the two bands. Then, the system completes the transition
from insulator to metal, known as the Wilson transition. Figure \ref{PhaseD}
is the phase diagram, being inversion invariant about the origin $%
(\kappa,\lambda)=(0,0)$. The dashed lines represent the topological phase
transition, and the solid black lines represent the Wilson transition. The
purple and cyan regions are the metallic and insulator phases, respectively;
being the topologically nontrivial phases. Similarly, the green and yellow
regions are the metallic and insulator phases, respectively; being the
trivial phases.

\begin{figure}[t]
\includegraphics[bb=0 0 580 800,width=8.8cm,clip]{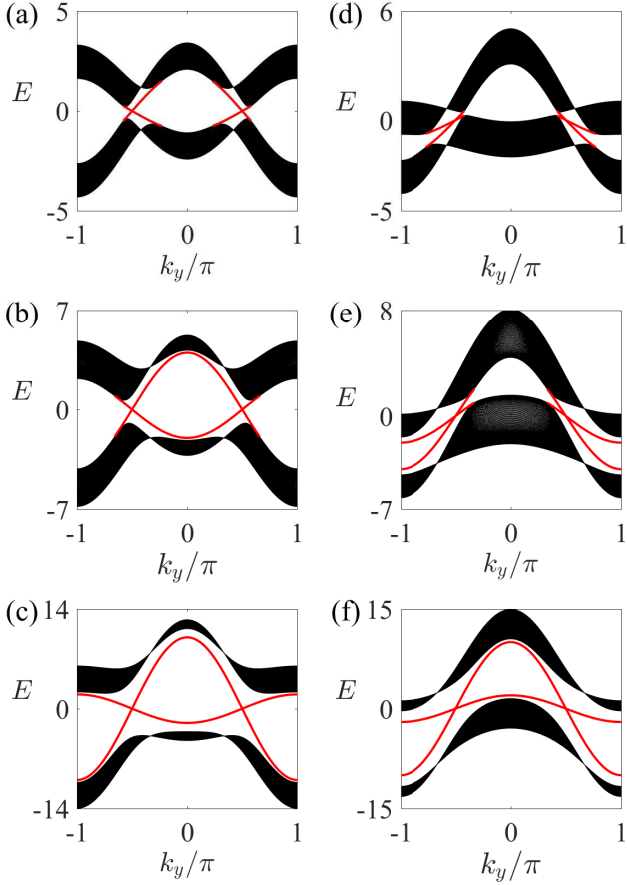}
\caption{Band structures of three cases in the insulator phase at (a) $%
\protect\kappa=1,\protect\lambda=-0.5$, (b) $\protect\kappa=2,\protect\lambda%
=-1$, (c) $\protect\kappa =5,\protect\lambda=-1 $ and three cases in the
metallic phase at (d) $\protect\kappa=1,\protect\lambda=0.5 $, (e) $\protect%
\kappa=2,\protect\lambda=1$, (f) $\protect\kappa=5,\protect\lambda=1$ as
indicated by the black solid dots in the phase diagram of Fig.~\protect\ref%
{PhaseD}. The energy band is depicted for the OBC in the $x$ direction and
the PBC in the $y$ direction. The other system parameter is $m=1$. The red
line indicates the edge state.}
\label{Topol}
\end{figure}

The topological insulator phase is indicated by the cyan region in the phase
diagram Fig. \ref{PhaseD}. The two bands are fully gapped as shown in the
energy spectra Figs. \ref{Topol}(a)-\ref{Topol}(c). The topological metallic
phase is indicated by the purple region in the phase diagram Fig. \ref%
{PhaseD}. The lowest energy of the conduction band is lower than the highest
energy of the valence band as shown in the energy spectra Figs. \ref{Topol}%
(d)-\ref{Topol}(f). When the system in the insulator phase crosses the solid
black lines in the phase diagram Fig. \ref{PhaseD}, the system undergoes a
Wilson phase transition and enters the metallic phase. However, the band gap
is always open, which ensures that the topological characteristics remain
unchanged. This indicates that the topological metallic phase and the
topological insulator phase have identical topological properties.

\begin{figure}[t]
\includegraphics[bb=0 0 580 800,width=8.8cm,clip]{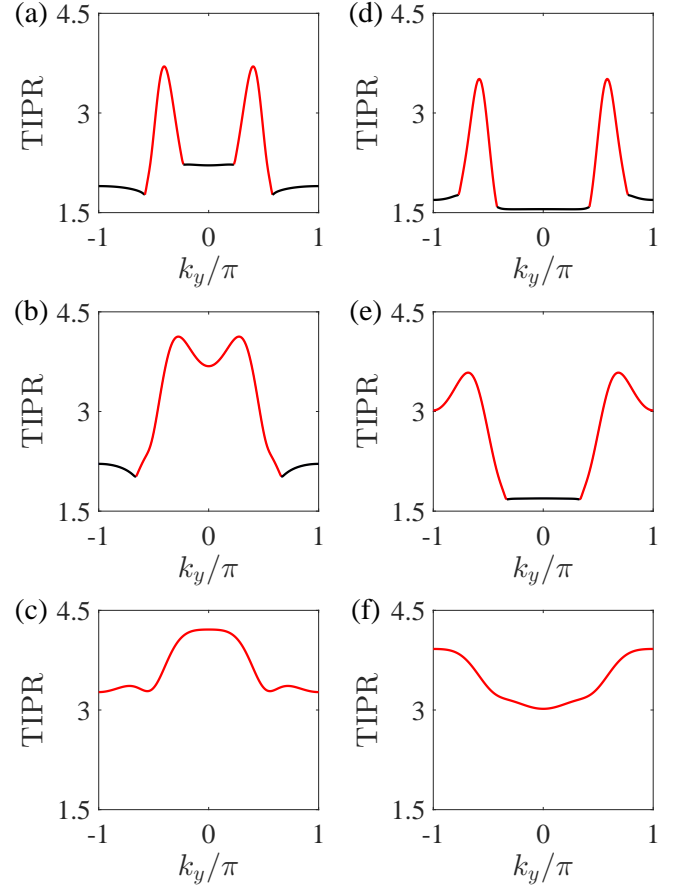}
\caption{Plot of the total inverse participation ratio along the momentum $%
k_{y}$ as the counterpart of Fig. \protect\ref{Topol}. The red line
represents the TIPR including the localized edge state, and the black line
represents the TIPR of extended states in bulk. (a) $\protect\kappa=1,%
\protect\lambda=-0.5$, (b) $\protect\kappa=2,\protect\lambda=-1$, (c) $%
\protect\kappa =5,\protect\lambda=-1$, (d) $\protect\kappa=1,\protect\lambda%
=0.5 $, (e) $\protect\kappa=2,\protect\lambda=1$, (f) $\protect\kappa=5,%
\protect\lambda=1$.}
\label{TIPR}
\end{figure}

\begin{figure}[t]
\includegraphics[bb=0 -20 585 800,width=8.8cm,clip]{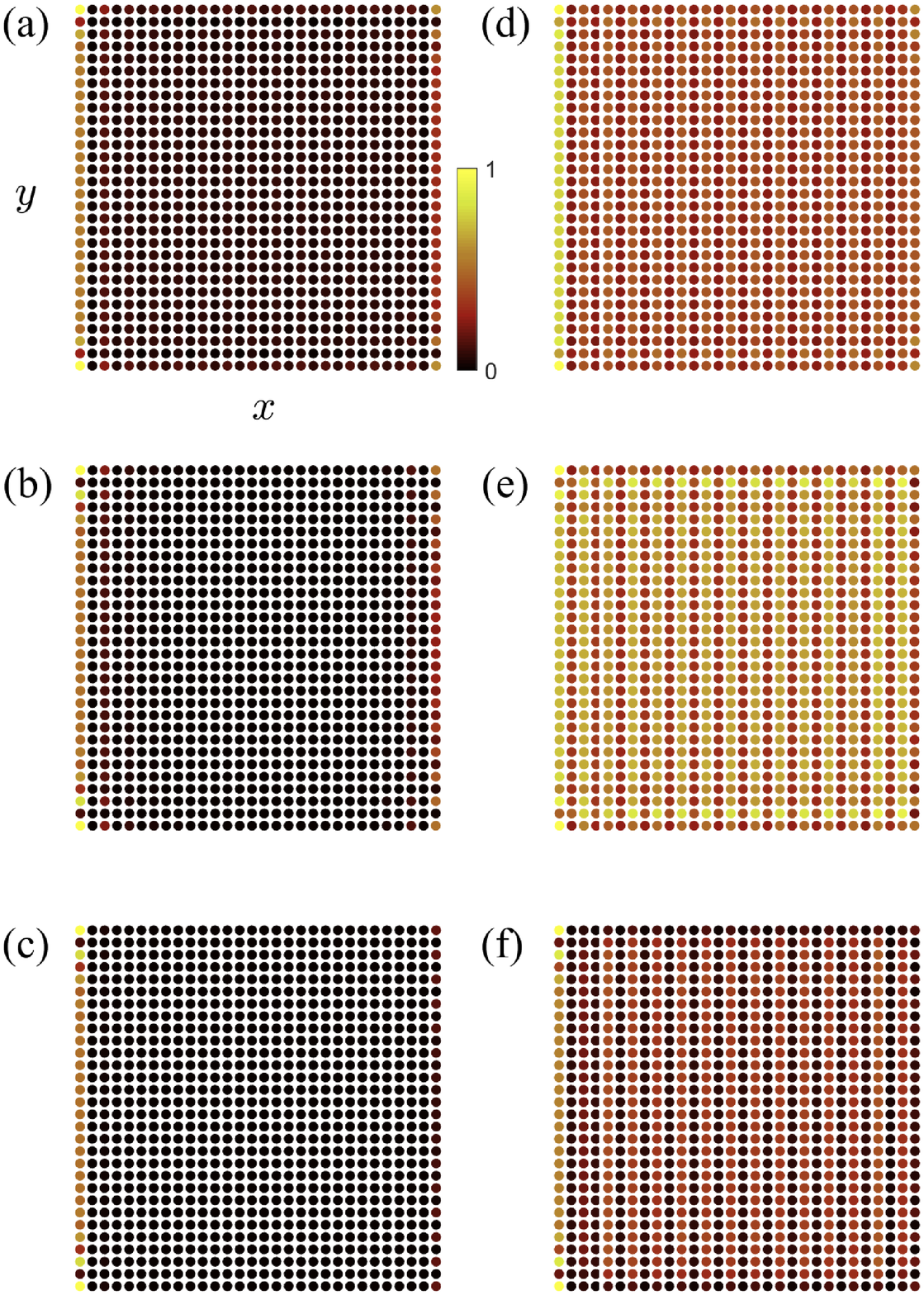}
\caption{Plot of the local density of state near the zero energy as the
counterpart of Fig. \protect\ref{Topol}. (a) $\protect\kappa=1,\protect%
\lambda=-0.5$, (b) $\protect\kappa=2,\protect\lambda=-1$, (c) $\protect%
\kappa =5,\protect\lambda=-1$, (d) $\protect\kappa=1,\protect\lambda=0.5 $,
(e) $\protect\kappa=2,\protect\lambda=1$, (f) $\protect\kappa=5,\protect%
\lambda=1$. Here we take $\protect\alpha=2$ and lattice size is $30\times30$%
. The color bar indicates the probability of wave function and the maximum
is renormalized to 1.}
\label{LDOS}
\end{figure}

Interestingly, the system has three different types of edge states featured
from the relation between the edge states and the bulk bands in both the
topological metallic phase and the topological insulator phase. The edge
states always intersect at the momentum $k_{y}=\pm \pi /2$, which is also
the band closing point. These are observed from the energy spectra of Figs. %
\ref{Topol}(a)-\ref{Topol}(c) and Figs. \ref{Topol}(d)-\ref{Topol}(f),
respectively. The different types of edge states can be deformed into each
other without the band gap closing; thus, they possess identical topological
properties. These edge states are distinguished by the lines $\lambda
+\kappa \pm 2=0$, $\lambda -\kappa \pm 2=0$ as the solid brown lines in the
phase diagram of Fig. \ref{PhaseD}. The topological insulator phase in the $%
\kappa >0$ is divided into three regions; the spectra for points $a$, $b$,
and $c$ are representative cases. In Fig. \ref{Topol}(a), the system has two
pairs of gapless edge states, connecting the upper and the lower bulk bands.
In Fig. \ref{Topol}(b), the gapless edge states near the center of the
Brillouin zone (BZ) detach the bulk bands, forming two edge states cross
twice at $k_{y}=\pm \pi /2$ with each edge state connected to one bulk band,
respectively. In Fig. \ref{Topol}(c), the edge states are completely
detached from the bulk bands and become the in-gap edge states. The edge
states in the topological metallic phase inherit the characteristics of the
edge states in the topological insulator phase. Figures \ref{Topol}(d) to %
\ref{Topol}(f) show the three typical energy spectra for the topological
metallic phase. The topological metal still has separable energy bands. In
Fig. \ref{Topol}(d), the gapless edge states intact the bulk bands. In Fig. %
\ref{Topol}(e), the edge states detach the bulk bands near the boundary of
the BZ. In Fig. \ref{Topol}(f), the edge states are completely detached from
the bulk band and become in-gap edge states.

For the three cases of insulator or metal as separated by the brown solid
lines in Fig.~\ref{PhaseD}, we introduce the total inverse participation
ratio (TIPR) as a criterion to distinguish their different features. The
existence of edge states can be measured by the quantity of TIPR, which is
defined as%
\begin{equation}
\text{TIPR}(k_{y})=\sum_{n,l}\left\vert \langle l\left\vert \psi
_{n,k_{y}}\right\rangle \right\vert ^{4},
\end{equation}%
where $n$ is the band index and $\left\vert \psi _{n,k_{y}}\right\rangle $
is eigenstate. In the large size limit, the value of TIPR$(k_{y})$ is
stable. The distribution of TIPR has three different configurations. The TIPR%
$(k_{y})$ has relative large values in two separated regions, in single
incomplete region, and in the whole region of the Brillouin zone of $k_{y}$.
The TIPR$(k_{y})$ in the topological insulator phase is shown in Figs.~\ref%
{TIPR}(a)-\ref{TIPR}(c) and the TIPR$(k_{y})$ in the topological metal phase
is shown in Figs.~\ref{TIPR}(d)-\ref{TIPR}(f). The distribution can be
characterized by the scattering process, where the mid-gap levels are
detected. Each region opens a window in the momentum space (or energy) for
the channel of electron transport, and then can be detected by measuring the
edge electrical conductivity and thermal conductivity \cite{XWen19}.

The insulator phase and metallic phase can be distinguished from the local
density of states (LDOS), which describes a space-resolved density of
states. The LDOS at the position $x$ and the energy $E$ is defined as $%
D(E,x)=\sum_{n}\left\vert \psi _{n}(x)\right\vert ^{2}\delta (E-\varepsilon
_{n})$ in a continuous system, and the LDOS reduces to the form 
\begin{equation}
D(E,l)=\frac{1}{\chi }\sum_{n=1}^{N}e^{-\alpha ^{2}\left( E_{n}-E\right)
^{2}}\left\vert \langle l\left\vert \psi _{n}\right\rangle \right\vert ^{2},
\label{DE}
\end{equation}%
at the lattice site ${l}$ and the energy ${E}$ in a discrete system. In Eq. (%
\ref{DE}), the delta-function is replaced by a Gaussian function and $\chi $
is the renormalization factor. Here, we consider the half-filling situation.
The LDOS at the Fermi surface $E_{F}=0$ is calculated for the finite size
systems with several representative parameters; the results are shown in
Fig.~\ref{LDOS}. $D(E,l)$ has vanishing distributions in all the insulating
phases as shown in Figs.~\ref{LDOS}(a)-\ref{LDOS}(c), but has finite
distributions in all the conducting phases as shown in Figs.~\ref{LDOS}(d)-%
\ref{LDOS}(f). In addition, ${D(E,l)}$ has evidently large distributions
along only two sides in all the phases, resulting the relatively large
electrical and thermal conductivities. The LDOS is detected by a scanning
tunneling microscope (STM), which is capable of imaging electron densities
of states with atomic resolution \cite{NHLe20}. The TIPR and LDOS are
measurable quantities in experiment. Furthermore, the nonzero bulk current
of the topological metal differs from that of the insulator and the
relatively larger edge current of the topological metal differs from that of
the normal conductor.

In the topological phases, the edge states exist for the system under the
periodic boundary condition (PBC) in the $y$ direction and the open boundary
condition (OBC) in the $x$ direction. The energies of the edge states that
localized on the left and right boundaries are 
\begin{equation}
E_{L}(k_{y})=2\lambda \cos k_{y},E_{R}(k_{y})=2\kappa \cos k_{y}.
\end{equation}

The wave function for the edge state with $E_{L}\left( k_{y}\right) $ [$%
E_{R}\left( k_{y}\right) $] localized at the left (right) boundary is
denoted as $\left\vert \psi _{L}\right\rangle $ ($\left\vert \psi
_{R}\right\rangle $). We denote the expression of the edge states as 
\begin{equation}
|\psi _{L,R}\rangle =\left( \psi _{1A},\psi _{1B},...,\psi _{NA},\psi
_{NB}\right) ,
\end{equation}%
where $N$ is the total number of the unit cells. The edge states at the
limitation of infinity large size system ($N\rightarrow \infty $) are
analytically obtained. For the edge state $|\psi _{L}\rangle $, the
components of $|\psi _{L}\rangle $ in the $n$-th unit cell are $(\psi
_{nA},\psi _{nB})=(\rho ^{n-1},0)$ with the decay factor $\rho =-(\lambda
+2\cos k_{y})/\kappa $. For the edge state $\left\vert \psi
_{R}\right\rangle $, the components of $\left\vert \psi _{R}\right\rangle $
in the $n$-th unit cell are $(\psi _{nA},\psi _{nB})=(0,\rho ^{N-n})$. The
expression is valid for all the three types of topological edge states.

\section{Topological characterization}

\label{IV}

Nonspatial (internal) symmetry plays a key role in the topological
classification and determines the topological characterization \cite%
{CKChiu16}. Time-reversal symmetry (TRS), particle-hole symmetry (PHS), and
chiral symmetry (CS) classify the topological phases in a ten-fold way \cite%
{SRu10}. The time-reversal symmetry is defined as 
\begin{equation}
\mathrm{TRS}: \mathcal{T}h\left( \mathbf{k}\right) \mathcal{T}^{-1}=h\left( -%
\mathbf{k}\right) ,\mathcal{T}^{2}=\pm 1.
\end{equation}%
The time-reversal operator $\mathcal{T}=UK$\ is an anti-unitary operator,
where $U$ is a linear operator and the complex conjugation $K$\ is an
anti-linear operator.

The TRS ensures that the Berry curvature $F$ satisfies $F(\mathbf{k})=-F(-%
\mathbf{k})$, the integral of $F$ in the entire BZ results in a vanishing
Chern number. How to characterize the band topology and determine the
topologically nontrivial phase of a time-reversal symmetric system? For the
time-reversal symmetric system with $\mathcal{T}^{2}=-1$, the topological
properties can be characterized by a $Z_{2}$ topological invariant. $Z_{2}=0$
represents the trivial insulator and $Z_{2}=1$ represents the topological
insulator. The Pfaffian method,\ time-reversal polarization, and non-Abelian
Berry connection have been developed as the $Z_{2}$ invariant, but these are
unable to describe the topological systems with $\mathcal{T}^{2}=+1$. For
the time-reversal symmetric system with $\mathcal{T}^{2}=+1$, the 2D Zak
phase, also known as the wave polarization, is used for the topological
characterization. The wave polarization is quantized to $0$ or $1/2$ under
the symmetry protection and is used to describe the band topology of the
time-reversal symmetric system with $\mathcal{T}^{2}=+1$ \cite{FLiu17}. The
topological edge states appear inside the band gap \cite{HCWu20}.

For the Bloch Hamiltonian of the 2D square lattice $h\left( \mathbf{k}%
\right) $, the linear operator $U=\sigma _{0}$ is the $2\times 2$ identical
matrix; thus, the 2D square lattice possesses the time-reversal symmetry of $%
\mathcal{T}^{2}=+1$. The term $d_{0}\sigma _{0}$ in the Bloch Hamiltonian $%
h\left( \mathbf{k}\right) $ is crucial for the appearance of the metallic
phase; however, $d_{0}\sigma _{0}$\ only adjusts the band energy, but not
the eigenstates of $h\left( \mathbf{k}\right) $. The topology of $h\left( 
\mathbf{k}\right) $ only relates to the eigenstate and thus is determined
from $\mathbf{d\cdot \sigma }$. The eigenstate of $h\left( \mathbf{k}\right) 
$ is in the form of 
\begin{equation}
\left\vert \Psi _{\pm }(\mathbf{k})\right\rangle =\frac{1}{\Omega _{\pm }}%
\dbinom{d_{z}\pm d}{d_{x}+id_{y}},
\end{equation}%
where the subscript $+$ represents the upper band and the subscript $-$
represents the lower band; and $\Omega _{\pm }=\sqrt{2d(d\pm d_{z})}$ is the
normalization factor.

The wave polarization is defined as%
\begin{equation}
\mathbf{P=}\frac{1}{2\pi }\int_{BZ}\mathrm{Tr}[\mathbf{A}%
(k_{x},k_{y})]dk_{x}dk_{y},
\end{equation}%
where $\mathbf{A}(k_{x},k_{y})=\left\langle \Psi (\mathbf{k})\right\vert
i\nabla _{\mathbf{k}}\left\vert \Psi (\mathbf{k})\right\rangle $ is the
Berry connection. $\mathbf{P=(}P_{x},P_{y}\mathbf{)}$ includes two
components, where $P_{x}$ characterizes the topology in the $x$ direction
and predicts the edge states for the 2D square lattice under the OBC in the $%
x$ direction. $P_{y}$ characterizes the topology in the $y$ direction. The
wave polarization defines the 2D Zak phase $\Theta _{\nu }=2\pi P_{\nu }$, ($%
\nu =x,y$). Protected by the reflection symmetry or inversion symmetry, the
value of wave polarization exhibits quantized, being $0$ or $\pm 1/2$. For
the Bloch Hamiltonian $h(\mathbf{k})$, the reflection symmetry in the $y$
direction guarantees $P_{y}=0$. The Bloch Hamiltonian $h(\mathbf{k})$ is
reflection symmetric in the $y$ direction $h(k_{x},k_{y})=h(k_{x},-k_{y})$. $%
P_{y}=0$ reflects the lack of nontrivial edge state for the 2D square
lattice under the OBC in the $y$ direction and the PBC in the $x$ direction.

\begin{figure}[t]
\includegraphics[bb=0 0 500 510,width=7.5 cm,clip]{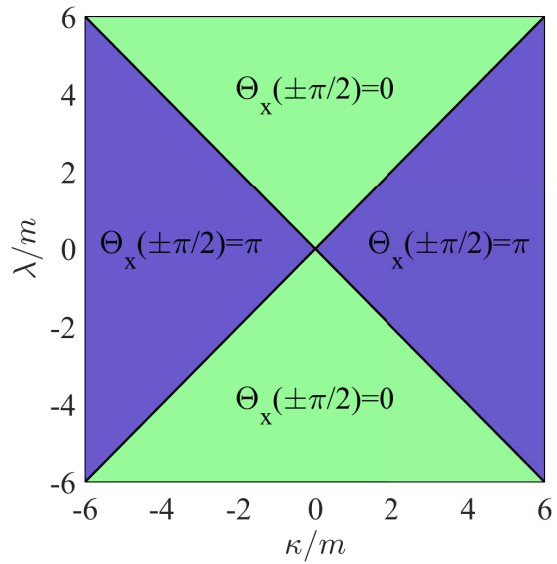}
\caption{Topological phase diagram of the time-reversal symmetric 2D square
lattice. The $|\protect\lambda |<|\protect\kappa |$ region in purple is the
topologically nontrivial phase; in contrast, the $|\protect\lambda |>|%
\protect\kappa |$ region in green is the trivial phase. The Zak phase $%
\Theta _{x}$ distinguishes different topological phases.}
\label{TPD}
\end{figure}

The polarization $P_{x}$ is not quantized without the symmetry protection.
Therefore, $P_{x}$ is invalid for the topological characterization. We apply
a Fourier transformation to decompose the 2D lattice Hamiltonian into
decoupled 1D lattice Hamiltonian parameterized by the momentum $k_{y}$. The
band gap closes at $k_{y}=\pm \pi /2$, where the topological phase
transition occurs and the edge states intersect. For $\lambda =-\kappa $ $%
(\kappa \neq 0)$, the band degenerate points locate at $(k_{x},k_{y})=(0,\pm
\pi /2)$. For $\lambda =\kappa $ $(|\kappa |\leq 1)$, the band gap closes at 
$(k_{x},k_{y})=(0,\pm \arccos \lambda )$ and $(\pi ,\pm \pi /2)$. For $%
\lambda =\kappa $ $(|\kappa |>1)$, the band degenerate points locate at $%
(k_{x},k_{y})=(\pi ,\pm \pi /2)$. In particular, when $(\lambda ,\kappa
)=(0,0)$, the band degenerate points become two degenerate lines at $%
k_{y}=\pm \pi /2$.

The 1D Zak phase is defined as 
\begin{equation}
\Theta _{x}\left( k_{y}\right) =\int \left\langle \psi (k_{x})\right\vert
i\partial _{k_{x}}\left\vert \psi (k_{x})\right\rangle dk_{x},
\end{equation}%
which reflects the winding of the Bloch Hamiltonian when $k_{x}$ varies
throughout the entire BZ of the decoupled 1D lattice. The 1D Zak phase $%
\Theta _{x}(\pm \pi /2)$ is quantized to $\pi $ in the topologically
nontrivial phases (the purple and cyan regions in Fig. \ref{PhaseD}),
whereas $\Theta _{x}(\pm \pi /2)$ is quantized to $0$ in the trivial phase
(the green and yellow regions in Fig. \ref{PhaseD}). The topological
characterization is shown in Fig. \ref{TPD} and is consistent with the phase
diagram in Fig. \ref{PhaseD}. $\Theta _{x}(\pm \pi /2)=\pi $ characterizes
the topological metal and topological insulator. $\Theta _{x}(\pm \pi /2)=0$
characterizes the ordinary metal and ordinary insulator.

\section{Discussion}

\label{V}

The vertical coupling $\lambda \neq \pm \kappa $ is crucial for the
topological metallic phase as indicated in Fig. \ref{PhaseD}. In addition,
we consider the horizontal couplings $\lambda $ and $\kappa $ are replaced
by $\lambda ^{\prime }$ and $\kappa ^{\prime }$. Then, when $\lambda
^{\prime }=\kappa ^{\prime }$, the energy band must have degenerate points
located at $(k_{x},k_{y})=(\pm \pi ,\pm \pi /2)$; when $\lambda ^{\prime
}=-\kappa ^{\prime }$, the energy band must have degenerate points located
at $(k_{x},k_{y})=(0,\pm \pi /2)$. Notably, $\lambda ^{\prime }\neq \pm
\kappa ^{\prime }$ is necessary for the topological metallic phase, but $%
\lambda =\lambda ^{\prime }$ and $\kappa =\kappa ^{\prime }$ are not
necessary. The nearest neighbor coupling $m\neq 0$ is important for the
different edge states in the topological metallic phases. For $m=0$, only
the in-gap edge states exist in the topological phase and only one type of
topological phase presents in this situation.

The spin-orbit coupling is used to create the magnetic flux and the
traditional topological metal. However, the proposed 2D topological metal is
time-reversal symmetric without the magnetic flux. This requires the
spin-orbit coupling is negligible in the material. The organic materials
composed of light elements have weak spin-orbit coupling \cite{LJin15}.
Therefore, the organic materials may be suitable candidates for the class AI
topological phase \cite{ZSLiao21}.

If the time-reversal symmetry breaks when the square lattice is threaded by
the magnetic flux. The topological metallic phase still exists through
tuning the magnetic flux and the coupling strengths. In this situation, the
topological phase is characterized by the nonzero Chern number. Under the
PBC in the $y$ direction, a pair of anti-chiral edge states exist after
imposing the OBC in the $x$ direction. The antichiral edge states also
connect the upper and lower energy bands to form the conductive surface
states, but the two antichiral edge states on different boundaries propagate
in the same direction in contrast to the chiral edge states \cite{EColo18}.

Furthermore, the non-Hermiticity compresses the band gap between the valence
band and the conduction band. The non-Hermitian topological metallic phase
exits in the proposed 2D square lattice when the additional gain and loss
are presented in different sublattices. This is an interesting topic that
deserving further investigation.

\section{Conclusion}

\label{VI}

In conclusion, we provide an alternative way of generating the topological
metals. The proposed topological metallic phases are protected by the
time-reversal symmetry and are free from magnetic flux. The time-reversal
symmetry ensures a zero Chern number; however, the 2D square lattice
supports topologically nontrivial phases, which are distinguished by the
different types of edge states. The 2D square lattice supports three
different cases in the topological metallic phase that significantly differ
from the traditional topological metallic phases created through tuning the
magnetic flux. These different cases can be deformed into each other without
topological phase transition. Our findings provide insights into the
time-reversal symmetric topological metals.

\section*{Acknowledgement}

This work was supported by the National Natural Science Foundation of China
(Grants No.~11975128 and No.~11874225).

\end{document}